\documentclass[journal = acsmacro,layout=twocolumn]{achemso}
\usepackage{dcolumn}
\usepackage{multicol}
\usepackage[dvipsnames]{xcolor}
\usepackage{soul}
\usepackage{bm}
\usepackage[utf8]{inputenc}
\usepackage{amsmath}
\usepackage{graphicx}
\usepackage{svg}
\usepackage{gensymb}
\usepackage{hyperref}

\def\DJ{{\fontencoding{T1}\selectfont\char208}}

\title{Near surface concentration profile of sheared semi-dilute polymer solutions}

\author{Suzanne Lafon}
 \email{slafon@phare.normalesup.org}
 \affiliation{Laboratoire de Physique des Solides, Université
Paris Saclay, CNRS, 91405 Orsay, France}
\author{Tiago Outerelo Corvo}
 \affiliation{Laboratoire de Physique des Solides, Université
Paris Saclay, CNRS, 91405 Orsay, France}
\author{Marion Grzelka}
 \affiliation{Laboratoire Léon Brillouin, CEA Saclay, 91191 Gif-sur-Yvette, France}
\author{Arnaud Hélary}
 \affiliation{Laboratoire Léon Brillouin, CEA Saclay, 91191 Gif-sur-Yvette, France}
\author{Philipp Gutfreund}
 \affiliation{Institut Laue-Langevin, 71 rue des Martyrs, 38000 Grenoble, France}
\author{Liliane Léger}
 \affiliation{Laboratoire de Physique des Solides, Université
Paris Saclay, CNRS, 91405 Orsay, France}
\author{Alexis Chennevière}
 \affiliation{Laboratoire Léon Brillouin, CEA Saclay, 91191 Gif-sur-Yvette, France}
\author{Frédéric Restagno}
 \affiliation{Laboratoire de Physique des Solides, Université
Paris Saclay, CNRS, 91405 Orsay, France}

\begin{document}

\begin{abstract}
Controlling the structure of polymer solutions near a solid surface is crucial for many industrial processes, as it significantly impacts solution flow and influences slip at the interface. To date, only a few techniques have been developed to experimentally investigate this type of interface at the nanometric scale of solid/liquid interaction. In this study, we probe the interface between a smooth sapphire surface and a semi-diluted polystyrene solution, using neutron reflectivity. A special setup for flow measurements under shear has been designed and optimized. Our results show that, at rest, polymer chains are globally depleted from the solid surface. Contrary to common assumptions, some polystyrene chains do adsorb onto the wall. Under flow conditions, we experimentally demonstrate that the depletion layer remains stable, a finding that has been hypothesized but only vaguely confirmed in the literature.

\end{abstract}


\section*{Introduction}
The interaction between a solid interface and polymer solutions has been and still is a subject of both theoretical and experimental studies\cite{stuart1985experimental, kawaguchi1992polymer, fleer1998general, degennesConformationsPolymersAttached1980}. It plays a major role in many processes, as polymers at interfaces can change stress transmission mechanisms by controlling slippage or providing targeted functionalities such as anti-fouling or bacteriostatic effects. The interaction between polymer chains and a solid interface can be either attractive, leading to adsorption of chains onto the surface or repulsive, leading to a so called depletion near interface. Several theoretical and experimental studies have highlighted the structure of absorbed polymer layers and its influence on the interfacial properties \cite{guiselin1992irreversible, thees2020review, napolitano2020}. The case of polymer depleted interfaces gave rise to a smaller amount of studies as the majority of neutral polymer / solid interfaces present adsorption. Some experimental measurements have indirectly measured the presence of a depletion layer when a semi-dilute polymer solution is placed onto a repulsive solid surface. These measurements are based on the increase of the fluid mobility close to the surface. Among these experiments, we can cite direct pressure drop measurements in porous media \cite{sorbie_depleted_1990,chauveteau_concentration_1984,zitha_unsteady-state_2001}, microfluidics experiments \cite{cuenca_submicron_2013,guyard2021near}, surface forces apparatus \cite{horn_hydrodynamic_2000,barraudLargeSlippageDepletion2019}, atomic fore microscopy \cite{borkovec2016,klitzing2020} or rheology \cite{sanchez-reyes_interfacial_2003,boukany_molecular_2010}.
Generally, directly measuring the concentration of polymer close to the interface is difficult. The first observation of depletion of a polymer solution near a solid surface has been made by Allain \textit{et al.} in 1982 \cite{allainDirectOpticalObservation1982} with evanescent waves. Since then, different techniques have been used to directly measure depletion, such as neutron reflectivity \cite{leeDirectMeasurementsPolymer1991} and optical reflectivity \cite{gvaramiaDepletionPolyelectrolytesLikeCharged2022}. Finally, the effect of flow on depletion is poorly studied because of the difficulty of the experiments, but for now, what is reported is a thickening of the depletion layer at high flow rates \cite{ausserreHydrodynamicThickeningDepletion1991, depabloHydrodynamicChangesDepletion1992} for rigid polymers.

In this paper, we use neutron reflectivity \cite{cousin2018neutron, penfold1990application, russell1990x} to directly measure depletion near a smooth sapphire surface for non-charged semi-dilute polystyrene solutions at rest and under shear thanks to the development of a custom flow cell. 

\section{Experimental methods}

\paragraph{Solutions} The solutions are made of fully deuterated polystyrene (dPS) (Polymer Source, $\text{\DJ} = 1.17$, $M_{n} = 195$ kg/mol or $M_{n} = 1.56$ Mg/mol) in hydrogenated diethylphtalate, DEP (Sigma Aldrich), which is a good solvent for PS at room temperature\cite{grzelka2021}. The volume fractions $\phi$ ($3$ and $6$ $\%$) are chosen to be in the semi-dilute regime $\phi > \phi^{*}$ with $\phi^{*} \approx N^{-4/5}$ the overlap volume fraction, which is about $0.24\, \%$ for the $195$ kg/mol polystyrene, and $0.06$ $\%$ for the $1.56$ Mg/mol one.
 
Solutions are homogenized under gentle stirring during two weeks prior to the experiments. The wall of interest is a polished sapphire surface from Fichou. Before experiment, the sapphire substrate is cleaned under a UV-ozone lamp (ProCleaner$^{\text{TM}}$ \textit{Plus}, Bioforce Nanosciences) for at least $30$ minutes.

\paragraph{Small angle neutron scattering}
SANS experiments are made on the spectrometer PACE in the Laboratoire Léon Brillouin
(CEA Saclay). By changing the distance $D$ between the sample and the detector and the
selected wavelength $\lambda$, it is possible to tune the studied range of scattering vector Q. Here, we used 4 configurations ($D$, $\lambda$): (1 m, 5 \AA), (3 m, 5~ \AA), (5 m, 5 \AA) and (5 m, 13 \AA), leading to the total $Q$ range of 0.04 to 0.4 \AA$^{-1}$. Standard corrections were applied for sample volume,
neutron beam transmission, empty cell signal, and detector efficiency to the raw signal to
obtain scattering spectra in absolute units. The blob size of the semi-dilute dPS/DEP solutions was extracted by fitting the scattering spectra by a Lorentzian function (see SI Fig.~S1)

\paragraph{Neutron reflectivity under shear}

\begin{figure}
	\includegraphics[width=\columnwidth]{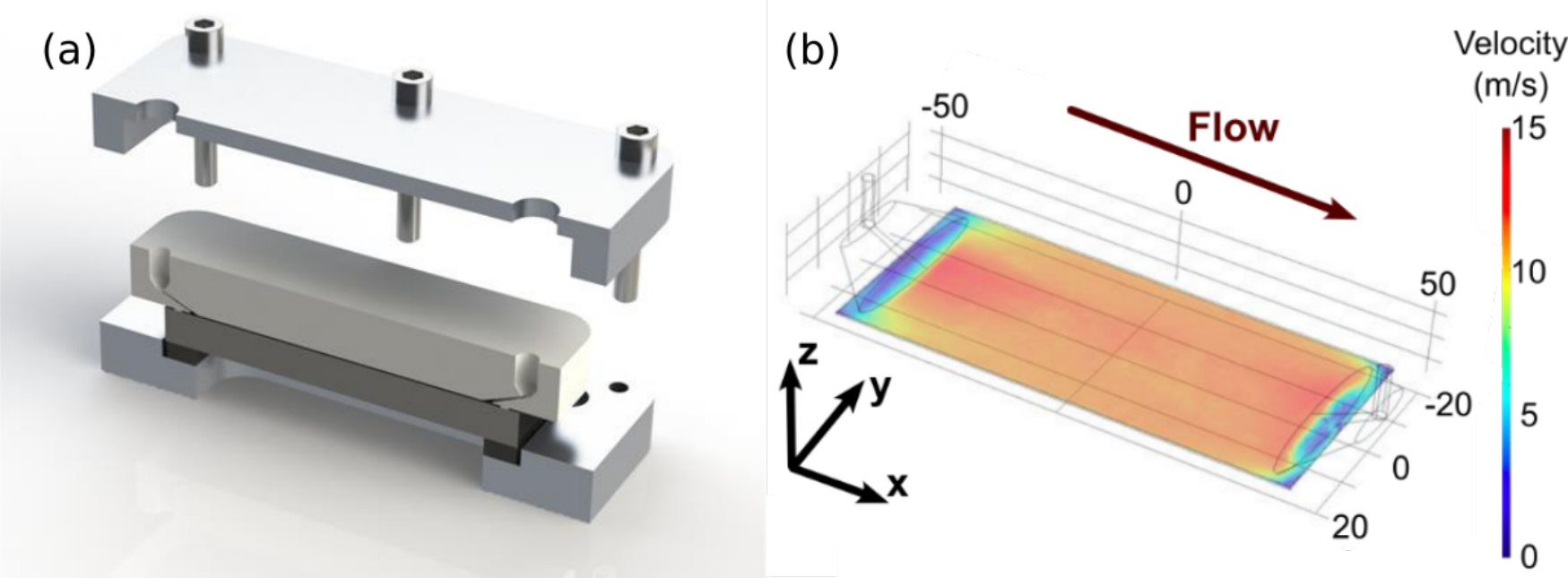}
	\caption{(a): 3D view of the shear cell used for neutron reflectivity. The fluid is injected through the PTFE block (white) block and flows on the substrate represented in dark gray. The fluid gap is 1~mm thick.\\(b): COMSOL flow simulation of a newtonian fluid ($\eta=11$~Pa.s)\label{fig:cell} }
\end{figure}

A custom Poiseuille flow cell designed to shear the polymer solution is depicted in Figure \ref{fig:cell}. The bottom surface is the sapphire substrate (4 x 10 x 2 cm), the top surface is made of a PTFE block. This block is machined with inlet nozzles which geometry allows to obtain a homogeneous shear rate over the illuminated surface. The height of the sandwiched liquid is controlled by a rectangular Viton frame of thickness $h ~=~ 1$ mm. The resulting sapphire/liquid/PTFE block is sandwiched between two large blocks of aluminium. The cell is then sealed by forcefully screwing blocks together. The flow is applied with a Chemyx Fusion 6000 syringe pump with 50 mL steel syringes, which flow rate can be varied between $10^{-4}$ and $270$ mL/min, both in injection and withdrawal.

Measurements are done either with a constant flow or with an alternating injection/withdrawal flow, in which case the acquisition of reflected neutrons is synchronized with the alternation frequency of the flow. We call the latter procedure "stroboscopic measurement". The experiment has been conducted on D17 \cite{saerbeck2018} neutron reflectometer at Institut Laue-Langevin (ILL). We use Time-Of-Flight (TOF) reflectivity at two angles of incidence $0.5 \degree$ and $2.5 \degree$, with acquisition times of $30$ and $60$ minutes respectively. The Scattering Length Densities (SLD) of the substrate and the solvent have been measured beforehand and are respectively $5.77\times10^{-6}$~\AA$^{-2}$ and $1.62\times10^{-6}$~\AA$^{-2}$  (see SI Fig.~S3). The surface roughness of the sapphire substrate was found to be 2.4 \AA.

\paragraph{X-ray reflectivity} The X-ray reflectivity (XRR) measurements were performed on a Xeuss 2.0 instrument (Xenocs) with a Cu k-$\alpha$ source of wavelength $1.54$ \r{A} and a Pilatus 1M 2D detector (Dectris). The experiment is conducted under vacuum, with a sample-detector distance of $1.214$ m. We use two collimation slits set to $0.5$ mm x $1$ mm and $0.3$ mm x $1$ mm (height x width).

\paragraph{Rheology of PS/DEP solutions}
The rheological properties of dPS DEP solutions was characterized using shear oscillatory rheology on an Antoon Paar MCR302 rheometer with sand-blasted cone plane geometry (25~mm radius, 2° cone angle). The experiment was conducted at 20~°C and the strain was set to $\gamma=7\%$ to be in the linear regime. The reptation time of the polymer solution was evaluated using the crossover between the storage and loss modulus (see SI Fig.~S5).

\section{Results and Discussion}
\subsubsection{No flow}

\begin{figure*}
    \centering
    \includegraphics[width=7in]{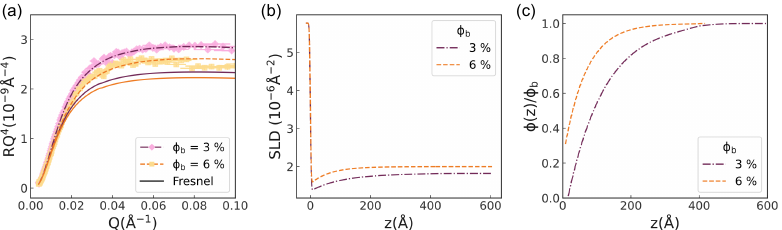}
    \caption{Effect of bulk volume fraction $\phi_{b}$. (a): reflectivity curves. Solid lines correspond to the Fresnel reflectivity if the solution was homogeneous until the solid surface. Dashed lines correspond to fits from which the SLD profiles (b) and thus the volume fraction profiles (c) are extracted.}
       \label{fig:phivar}
\end{figure*}

As a first result, Figure~\ref{fig:phivar} (left) shows the reflectivity curves of two static solutions at different bulk volume fractions $\phi_{\text{b}}$
of polymers, with the same molar mass $M_{n} = 195$ kg/mol. The data are plotted in the Porod representation ($RQ^4$ vs $Q$) where $Q$ is the scattering vector and $R$ is the reflectivity signal, to enhance potential differences between the two curves. For both volume fractions, the computed Fresnel reflectivity \textendash \, which is the reflectivity of an ideally smooth interface between two homogenous domains \cite{cousin2018neutron} \textendash \, is plotted with solid lines.

The observed differences between the two experimental curves can stem from the difference in coherent neutron scattering length density of the solutions and/or a change in the polymer segment density profile. As shown in figure~\ref{fig:phivar}, the computed Fresnel reflectivity does not allow to describe the experimental data, which highlights that near surface polymer concentration is not equal to the bulk concentration. In order to get a quantitative description of the interfacial structure, the data were fitted using the refnx Python module \cite{nelson2019refnx} assuming an exponential evolution of the volume fraction profile $\phi(z)$ as proposed by de Gennes \cite{degennesPolymerSolutionsInterface1981}: 

\begin{equation}
    \phi(z) =   \phi_\text{b} + (\phi_\text{w} - \phi_\text{b}) \text{e}^{-z/d} 
\end{equation}
where $z$ is the distance from the sapphire surface, $\phi_{\text{w}}$ and $\phi_\text{b}$ are the surface and bulk volume fractions, respectively, and $d$ is the typical size over which the polymer concentration differs from the bulk one. $\phi_\text{w}>\phi_\text{b}$ corresponds to adsorption while $\phi_\text{w}<\phi_\text{b}$ corresponds to depletion.

\begin{figure}
	\includegraphics[width=8.6cm, trim={0cm 0cm 0.0cm 0cm}]{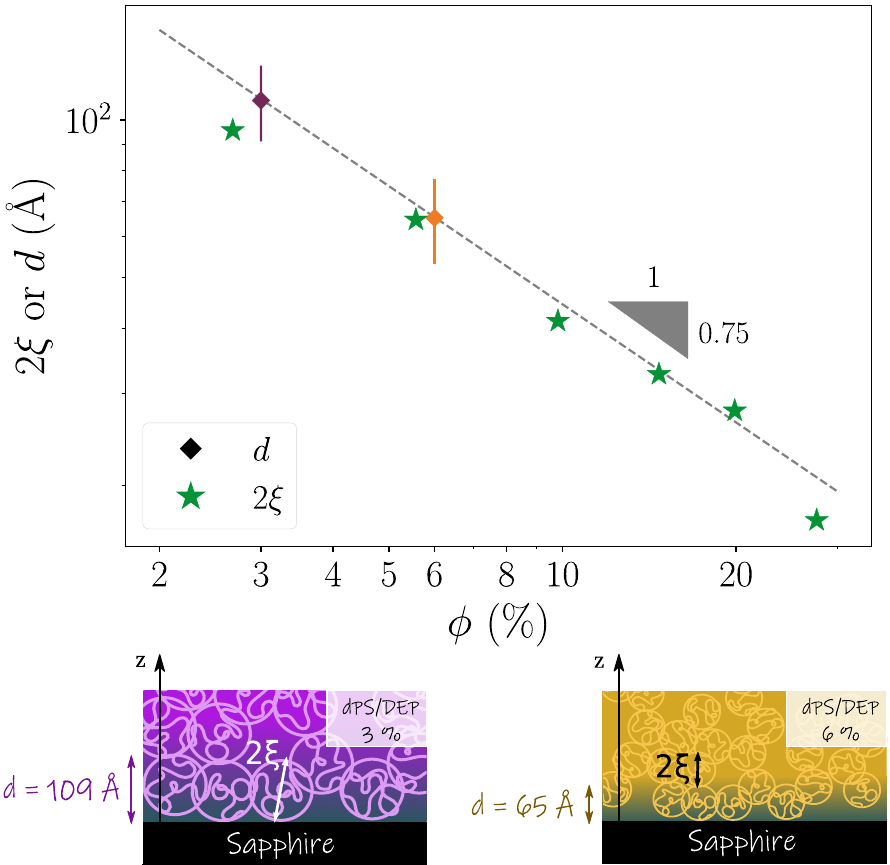}
	\caption{Top: Depletion layer thickness $d$ issued from the fit of the reflectivity data and blob size of the bulk solution measured using SANS as a function of the polymer volume fraction. The dashed line correspond to the scaling law of the blob size in good solvent conditions ($\xi\approx\phi^{-3/4}$). Bottom: Cartoon of the interface. The size of the depletion layer $d$ is typically the diameter of the blob $2 \xi$}.\label{fig:xi_d_phi}
\end{figure}

The resulting neutron scattering length density (SLD) profiles and polymer segment density profiles are shown in Figure \ref{fig:phivar}. At $z = 0$, the SLD decreases sharply because of the transition between the high-SLD smooth sapphire and the low-SLD solution. Far from the surface, the SLD reaches the bulk SLD of the solution. In the vicinity of the interface, data fitting shows unambiguously a polymer depletion which characteristic decay length $d$ decreases from $d=109 \pm 18$ \r{A} for $\phi = 3~\%$ to  $d= 65 \pm 12$ \r{A} at $\phi = 6~\%$. Previous neutron reflectivity measurements at the air/polymer solution interfaces was fitted with hyperbolic tangent function as predicted from the mean field theory \cite{joanny1979, leeDirectMeasurementsPolymer1991} but in the present work, exponential function ended with a better fit which may be due to the fact that the current substrate is not purely repulsive. We plotted in Figure \ref{fig:xi_d_phi} the evolution of both the depletion distance $d$ and the bulk blob size, $\xi$, measured using Small Angle Neutron Scattering. It can be observed that the characteristic depletion distance $d$ is comparable to $\xi$ and is in agreement with the bulk scaling law $\xi\propto \phi^{-3/4}$ \cite{rubinsteinPolymerPhysics}. 

This is in good agreement with theoretical predictions which state that the size of the depletion layer is the blob size \cite{silberberg1968,joanny1979,degennesPolymerSolutionsInterface1981} and consistent with previous measurements at liquid/air interface \cite{leeDirectMeasurementsPolymer1991}. This is schematically illustrated in Figure \ref{fig:xi_d_phi}. It must be pointed out here the concentration range accessible in this present experiment is much lower than in bulk. This stems from the fact that a low concentration implies a larger depletion distance but a much lower contrast between the depleted layer and the bulk. On the other hand, large concentrations can increase the contrast but decrease the size of the depleted layer. The deviation from the Fresnel reflectivity signal due to the presence of the depleted layer will appear at larger scattering vector value where the reflectivity signal to noise ratio is much lower. The two bulk concentrations used in the present work are a good comprise to characterize, as accurately as possible, the depletion at the interface.

\begin{figure}[!h]
    \centering
    \includegraphics[width=3.33in]{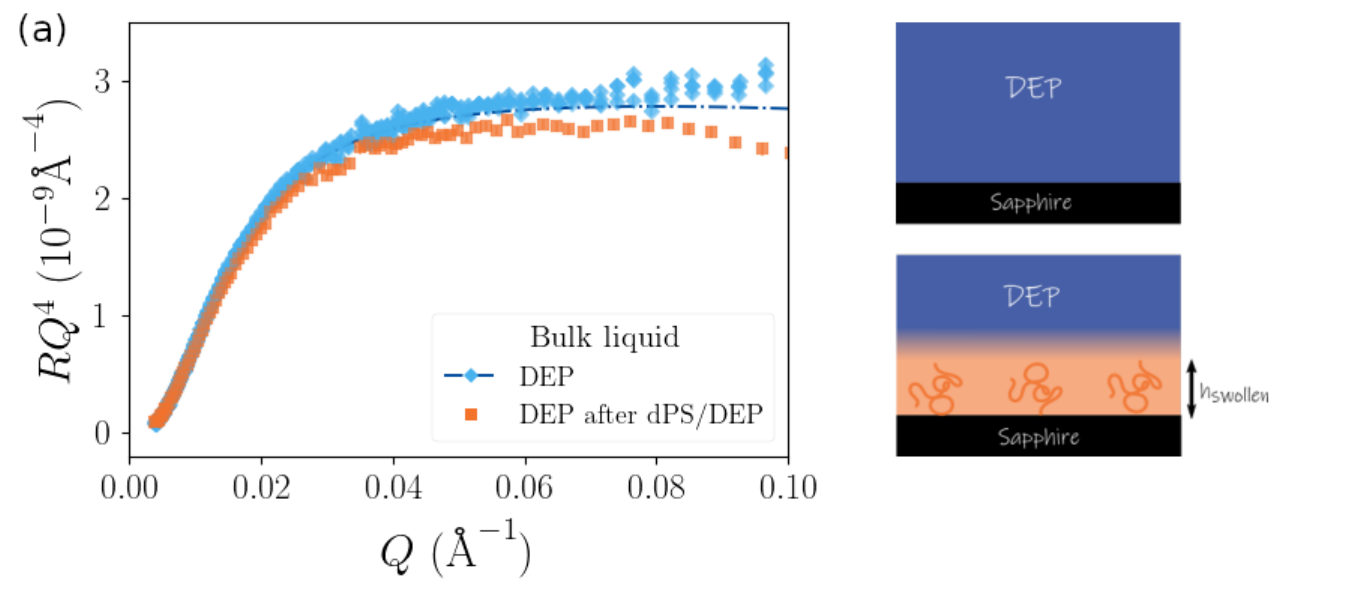}
    \includegraphics[width=3.33in]{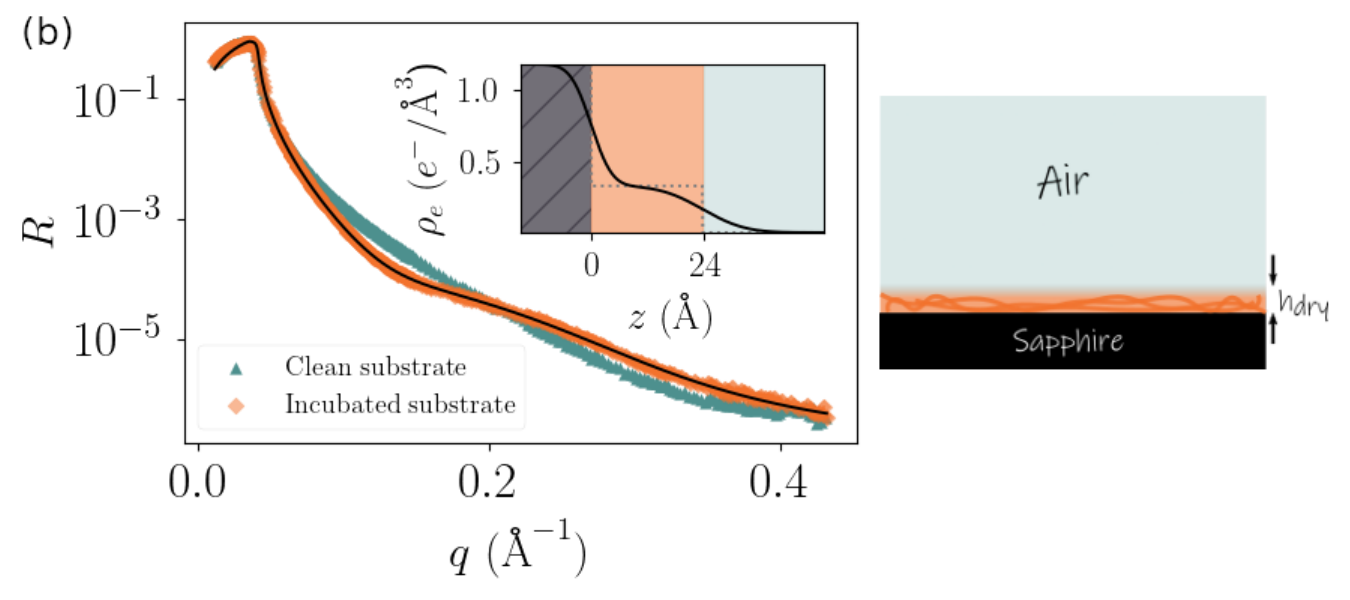}
    \caption{(a): Neutron reflectivity curves of DEP on clean sapphire (blue diamonds, fit is the blue dashed line) and on sapphire which has been incubated with dPS/DEP solution prior to the experiment (orange squares). (b): X-ray reflectivity curve of the sapphire surface that has been incubated with PS/DEP solution ($\phi = 6 \%$, $M_{n} = 708$ kg/mol), and then rinsed and dried (orange points). The black line is a fit with a rough interfacial layer. Green points correspond to the reflectivity of the clean substrate. Inset: Electronic density as a function of the distance from the interface. Corresponding cartoons are plotted on the right.}
    \label{fig:ads-depl}
\end{figure}

The observed depletion layer leads to the conclusion that polystyrene chains are less attracted by the sapphire compared to the solvent.
In Figure \ref{fig:phivar}, one can observe that the polymer volume fraction at the interface can be larger than zero. It means that some monomers are in contact with the surface but the lone segment density profile does not allow to conclude if they are physically adsorbed. In order to probe adsorption, we have compared the reflectivity curves of the solvent either on a clean sapphire and on a sapphire that has previously been in contact with the dPS/DEP solution ($\phi = 6$ \%, $M_{n} = 1.56$ Mg/mol). The resulting neutron reflectivity (NR) curves are plotted in Figure~\ref{fig:ads-depl} (top). One can see that at large $Q$ values, the profiles are significantly different. If there was pure depletion of dPS from the interface, the two profiles would be the same. It must be pointed out here that the surface is only thoroughly rinsed in DEP without any drying step in order to let the substrate transit from a contact with a binary solution (dPS/DEP) to only pure solvent without passing by a step where the substrate could be only in contact with PS and thus changing the interfacial energy of the system close to what can be observed in polymer thin films. Our measurement suggests that some dPS chains remain adsorbed on the interface. However, due to the low contrast between the adsorbed chains and the solvent, it is impossible to extract quantitative information from this measurement. 

To confirm this hypothesis, we have conducted X-ray reflectivity on an air/sapphire interface for a sapphire that had been in contact with the polystyrene solution for one hour, then thoroughly rinsed with DEP and dried. The corresponding reflectivity curve is plotted in Figure~\ref{fig:ads-depl} (bottom) and shows a clear Kiessig fringe between $Q=0.15$ \AA$^{-1}$ and $Q~=~0.3~$ \AA$^{-1}$. A single layer model with roughness at both interfaces fits well the data and confirms the presence of a dry adsorbed layer of thickness $h_\text{dry} = 24$ \r{A}. This value allows to estimate the surface density of chains $\nu = h_\text{dry} \frac{\rho N_{A}}{M_{n}} \approx 0.002$ nm$^{-2}$ with $\rho = 1040$ kg.m$^{-3}$ the density of polystyrene. Assuming that we have an Alexander - de Gennes brush\cite{degennesConformationsPolymersAttached1980}, we can estimate the mean volume fraction inside the swollen layer of adsorbed chains $\bar{\phi} \equiv \frac{Na^{3}}{D^{2}L}$ with $N\approx 6808$ the number of monomers per chain and $a\approx 5.5~\text{ \r{A}}$ the monomer size (estimated using $a^3 = \frac{M_\text{0}}{\rho N_{A}}$, with $M_\text{0}=104$ g/mol the molar mass of the monomer). $D = \frac{1}{\sqrt{\nu}} ~\approx~ 22~$ nm is the distance between adsorbed chains and $L ~=~ aN (\frac{a}{D})^{2/3} ~\approx ~320$ nm is the thickness of the swollen adsorbed layer in a good solvent (typically DEP). We find that $\bar{\phi} \approx 0.7$ \%. Fitting neutron reflectivity data from  such a thick swollen layer with a low amount of polymers is indeed very challenging.

It is quite surprising to see both depletion and adsorption as they are usually excluding scenari. The measurement of depletion is rather robust since the neutron flux of the ILL is strong enough to give precise measurements, and we could not fit any adsorption profile on our data while depletion profiles were easily fitted with good $\Tilde{\chi}^{2}$ values (between $0.8$ and $3.3$). As for adsorption, both the qualitative NR measurements and the quantitative X-Ray ones leave little doubts on the presence of remaining chains on the surface. In addition, Barraud \textit{et al.} \cite{barraudLargeSlippageDepletion2019} have mentioned an adsorption/depletion situation for a charged polymer on a metallic surface: they observed a depletion layer of the polyelectrolyte above its own adsorbed layer. However, in their case, adsorption was the result of the favorable electrostatic attraction between the metallic surface and the chains, and depletion was the consequence of the electrostatic repulsion between the adsorbed layer and the bulk chains. These arguments do not apply to our non-charged polymer.

From a chemical point of view, the surface of the sapphire displays hydroxyl groups with a surface density in the range of $1-10$ OH/nm$^{2}$ \cite{shenSurfactinStructuresInterfaces2011,merleHighFieldNMR2022}. Contrary to PS (hydrogenated or deuterated), DEP is able to make hydrogen bonds with these exposed OH groups, which is likely to favor the DEP-surface interaction compared to the dPS-surface interaction which further corroborates the depletion scenario. However, adsorption of PS chains onto sapphire surfaces has already been mentionned in the literature for PS melts \cite{Tatek2011}, which means that PS can also interact favorably with the surface. Indeed, PS chains are able to adsorb onto exposed hydroxyl groups through $\pi$-H interaction with their phenyl groups \cite{meyerInteractionsAromaticRings2003, paridaAdsorptionOrganicMolecules2006}. The fact that we globally see depletion suggests that this interaction is weaker than the H-bonds between DEP and hydroxyl groups, but it might not exclude some PS chains to still adsorb. Another explanation of the adsorption of PS on the surface might be a potential inhomogeneity of the roughness on the sapphire surface: it has indeed been shown \cite{napolitano-asperities-2020} that surface asperities can notably modify the chain dynamics, even at relatively low roughness. Our experiment do not provide information on the in-plane structure of the substrate, and therefore we cannot discriminate between a chemical or a physical origin of the adsorption of PS on the sapphire surface.

\subsubsection{Under flow}

\begin{figure}
    \centering
    \includegraphics[width=3.33in]{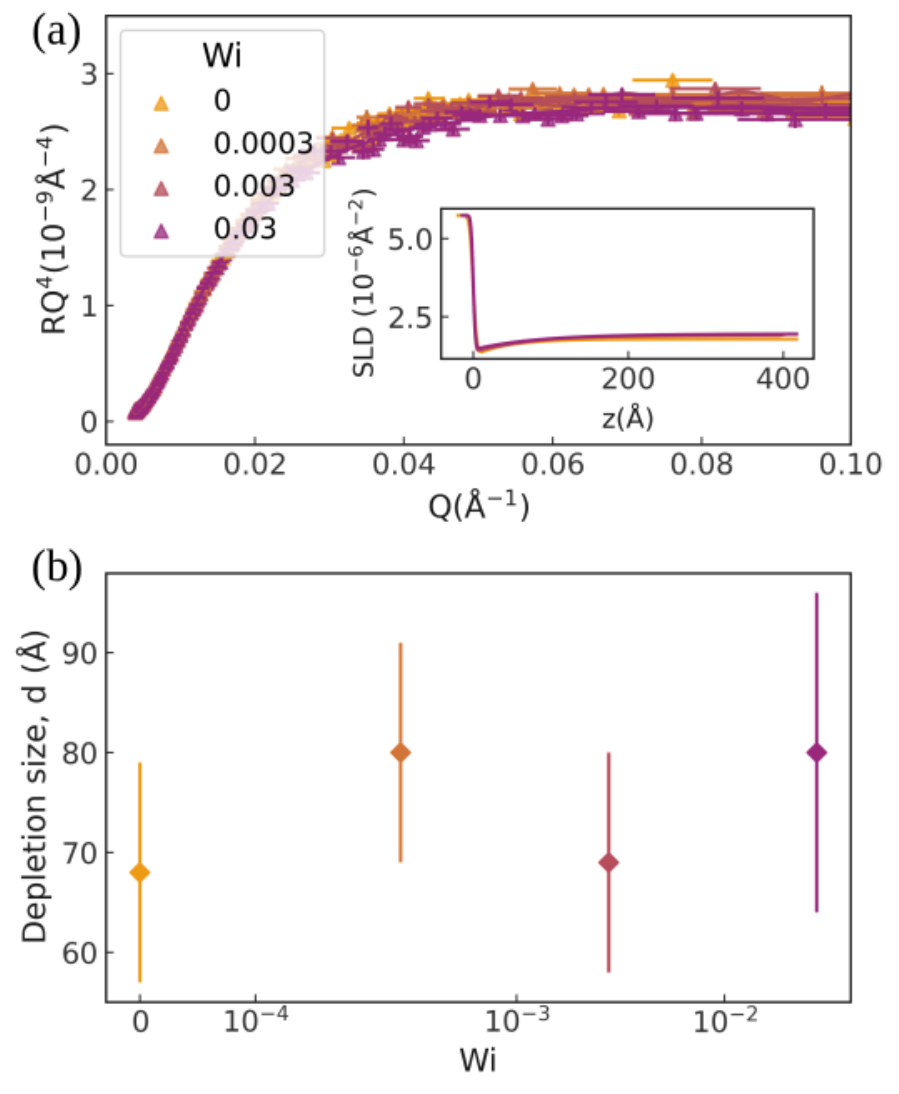}
    \caption{Effect of the flow on the reflectivity profiles. (a): reflectivity curves and corresponding SLD profiles in the inset. (b): Size of the depletion layer extracted from the fits as a function of the Weissenberg number Wi. These measurements have been done with $1.56$ Mg/mol dPS/DEP at a volume fraction $\phi = 6\: \%$.}
    \label{fig:Qvar}
\end{figure}

Finally, we have studied the effect of a Poiseuille flow on the concentration profiles of the dPS/DEP solutions near the sapphire surface. The results are plotted in Fig.~\ref{fig:Qvar}, for a solution with a $6~\%$ volume fraction and a $1.56$ Mg/mol molar mass. The flow rate is characterized by the Weissenberg number Wi, which is the dimensionless number comparing the typical relaxation time of the polymer solution ($\tau$) and the typical time scale of the flow ($\dot{\gamma}_{\text{max}}^{-1}$): Wi$ = \dot{\gamma}_{\text{max}} \tau$. In our geometry, the maximum shear rate $\dot{\gamma}_{\text{max}}$ is given by $6 Q_v /(\ell h^{2})$ with $Q_v$ the flow rate imposed by the pump, and $\ell$ and $h$ the width and the height of the cell, respectively. As for the typical relaxation time, we use the reptation time (also called terminal relaxation time) $\tau_{\text{rept}}$ of the solution. Oscillatory rheology (described in SI Fig.~S5) gave $\tau_{\text{rept}} = 0.23 \pm 0.01$ s.

The reflectivity curves are very similar for both the static solution and the flowing ones, and the fit yields the same SLD profiles, with $\tilde{\chi}^{2}$ between $1$ and $2$. The size of the depletion layer does not vary significantly with the Weissenberg number up to Wi $ = 0.03$. Reaching higher values of Wi is challenging because it requires both a highly viscous liquid (high $\tau$) and a strong flow rate $Q_v$, and we are limited by either the upper limit of flow rates accessible by the pump or by the total amount of solution we have, which conditions the emptying time of the syringe.

The flow rate can have an effect on both the depletion layer and on the adsorbed chains. Previous works on the effect of the flow rate on the size of the depletion layer report interesting behaviors. For rigid rodlike particles, Ausserré \textit{et al.} have shown that the depletion size $d$ increases with the flow rate due to hydrodynamics lift \cite{ausserreHydrodynamicThickeningDepletion1991}. For dilute polymer solutions, de Pablo \textit{et al.} have predicted a decreasing $d$ with the flow rate at moderate flow rates and an increasing $d$ at large flow rates \cite{depabloHydrodynamicChangesDepletion1992}, which corroborates the experiment done by Ausserré \textit{et al.}. They have used a dumbbell model, which aligns with the flow at low flow rates and starts to rotate quickly at high flow rates, so that the volume occupied by the dumbbell is larger and thus it flows further away from the wall. On the contrary, in our experiments, the dispersed polymers are flexible and therefore this description does not apply to our system. Furthermore, the presence of surface anchored chains that can entangle with free chains strongly changes the stress transmission mechanism at interface  \cite{morfin1999temperature,brochard1992shear,chenneviere_direct_2016}

Korolkovas \textit{et al.} \cite{korolkovas2017} have used neutron reflectivity to study the effect of flow rate on the interface between a dPS/DEP solution and PS brushes in a cone-plate rheometer. They have shown that the thickness of the brush decreases when increasing the Weissenberg number above $1$, which in their case is $\dot{\gamma} \tau$, with $\dot{\gamma}$ the applied shear rate. The grafting densities of their brushes varied between $0.04$ and $0.4$ nm$^{-2}$.
In our case, the polymer chains are just very weakly physiosorbed and cannot be distinguished from the bulk chains since they have the same neutron scattering length density. We thus really focus on the vicinity of the interface, the depletion layer in the present experiment. Even though surface anchored chains may endure a change in conformation, the local depletion layer does not evolve up to the maximum Weissenberg studied here. 

\section{Conclusion}

In conclusion, we have shown directly using neutron reflectivity that dPS/DEP solutions near a sapphire surface exhibits depletion. Interestingly, we show that depletion of polymer chains does not prevent some chains to adsorb onto the sapphire surface, probably due to a favorable interaction between the phenyl groups of the chains and the exposed hydroxyl groups of the solid surface. In addition, we show that the flow rate has no effect on the depletion layer up to Weissenberg number Wi $\approx 0.03$. These results provide valuable insight into the interfacial structure of a semi-dilute polymer solution flowing over smooth surfaces. Since depletion depends of solvent affinity with the surface, using good solvent of PS with different surface tension could be a suitable strategy to finely tune the depletion mechanism at interface. Besides, extending the range of flow rates to larger Weissenberg numbers might reveal a regime in which the depletion layer becomes shear-rate dependent. Indeed, in bulk, it was shown that large Weissenberg number can create concentration fluctuation at characteristic scales much larger than the blob size \cite{morfin1999temperature} but their influence on the local depletion layer remains up to know a very interesting experimental challenge broadening the scope of the current results.

\section{Supporting Information}
SANS measurement of the blob size (Figure S1). 3D drawing of the cell designed for the neutron reflectivity experiment under flow (Figure S2). Reflectivity curve of the pure DEP/sapphire interface (Figure S3). Summary of the Scattering Length Densities (SLD) of the various materials of the experiment (Figure S4). Shear rheology of the dPS/DEP solution (Figure S5).

\begin{acknowledgement}
This work was supported by the ANR-POILLU program (Grant No. ANR-19-CE06-007) and by the ANR-Shear@Interface (Grant No. ANR-22-CE92-0054). We thank O. Tessier for machining the cell. We thank the ILL for beamtime (DOI: 10.5291/ILL-DATA.9-11-2020).
\end{acknowledgement}

\bibliography{biblio.bib}

@article{allainDirectOpticalObservation1982,
  title = {Direct Optical Observation of Interfacial Depletion Layers in Polymer Solutions},
  author = {Allain, C. and Ausserr\'e, D. and Rondelez, F.},
  journal = {PRL},
  volume = {49},
  issue = {23},
  pages = {1694--1697},
  numpages = {0},
  year = {1982},
  month = {Dec},
  publisher = {American Physical Society},
  doi = {10.1103/PhysRevLett.49.1694},
  url = {https://link.aps.org/doi/10.1103/PhysRevLett.49.1694}
}

@article{napolitano-asperities-2020,
	title = {Substrate Roughness Speeds Up Segmental Dynamics of Thin Polymer Films},
	journal = {PRL},
	author = {Panagopoulou, Anna and Rodriguez-Tinoco, Cristian and White, Ronald P. and Lipson, Jane E. G. and Napolitnano, Simone},
	year = {2020},
    volume = {124},
    pages = {027802},
    issue = {027802},
    month = {Jan}

}

@article{chenneviere_direct_2016,
	title = {Direct {Molecular} {Evidence} of the {Origin} of {Slip} of {Polymer} {Melts} on {Grafted} {Brushes}},
	volume = {49},
	doi = {10.1021/acs.macromol.5b02505},
	number = {6},
	journal = {Macromol},
	author = {Chennevière, A. and Cousin, F. and Boué, F. and Drockenmuller, E. and Shull, K.R. and Léger, L. and Restagno, F.},
	year = {2016},
	pages = {2348--2353},
}

@article{klitzing2020,
	title = {Recent progress in measurements of oscillatory forces and liquid properties under confinement},
	volume = {47},
	doi = {10.1016/j.cocis.2020.02.002},
	journal = {Current Opinion in Colloied \& Interface Science},
	author = {Ludwig M. and von Klitzing R.},
	year = {2020},
	pages = {137-152},
}

@article{borkovec2016,
	title = {Interplay between Depletion and Double-Layer Forces Acting between Charged Particles in Solutions of Like-Charged Polyelectrolytes},
	volume = {117},
	doi = {10.1103/PhysRevLett.117.088001},
	pages = {088001},
	journal = {PRL},
	author = {Moazzami-Gudarz, M. and Kremer, T. and Valmacco, V. and Maroni, P. and Borkovec, M. and Trefalt, G.},
	year = {2016},
}

@article{sanchez-reyes_interfacial_2003,
	title = {Interfacial slip violations in polymer solutions: {Role} of microscale surface roughness},
	volume = {19},
	issn = {0743-7463},
	url = {<Go to ISI>://WOS:000182389100034},
	journal = {Langmuir},
	author = {Sanchez-Reyes, J. and Archer, L. A.},
	year = {2003},
	pages = {3304--3312--},
}

@article{boukany_molecular_2010,
	title = {Molecular {Imaging} of {Slip} in {Entangled} {DNA} {Solution}},
	issn = {0031-9007, 1079-7114},
	url = {http://link.aps.org/doi/10.1103/PhysRevLett.105.027802},
	doi = {10.1103/PhysRevLett.105.027802},
	volume = {105},
    issue = {2},
    pages = {027802},
	urldate = {2015-04-03},
	journal = {PRL},
	author = {Boukany, Pouyan E. and Hemminger, Orin and Wang, Shi-Qing and Lee, L. J.},
	year = {2010},
}

@article{horn_hydrodynamic_2000,
	title = {Hydrodynamic slippage inferred from thin film drainage measurements in a solution of nonadsorbing polymer},
	volume = {112},
	issn = {0021-9606},
	url = {http://aip.scitation.org/doi/abs/10.1063/1.481274},
	doi = {10.1063/1.481274},
	number = {14},
	urldate = {2023-03-28},
	journal = {J. Chem. Phys.},
	author = {Horn, Roger G. and Vinogradova, Olga I. and Mackay, Michael E. and Phan-Thien, Nhan},
	month = apr,
	year = {2000},
	note = {Publisher: American Institute of Physics},
	pages = {6424--6433},
}

@article{sorbie_depleted_1990,
	title = {Depleted {Layer} {Effects} in {Polymer} {Flow} through {Porous}  {Media}},
	volume = {139},
	url = {https://ac-els-cdn-com.inp.bib.cnrs.fr/002197979090103U/1-s2.0-002197979090103U-main.pdf?_tid=d69ba54e-ba55-11e7-aea7-00000aacb361&acdnat=1509026529_d9df106c844c8fe176bbc2ef22af3b57},
	urldate = {2017-10-26},
	journal = {J. Colloid Interface Sci.},
	author = {Sorbie, K. S.},
	year = {1990},
	pages = {299--314},
}

@article{zitha_unsteady-state_2001,
	title = {Unsteady-state flow of flexible polymers in porous media},
	volume = {234},
	number = {2},
	journal = {J. Colloid Interface Sci.},
	author = {Zitha, P. L. J. and Chauveteau, G. and Léger, L.},
	year = {2001},
	pages = {269--283},
}

@article{chauveteau_concentration_1984,
	title = {Concentration dependence of the effective viscosity of polymer solutions in small pores with repulsive or attractive walls},
	volume = {100},
	url = {http://www.sciencedirect.com/science/article/pii/0021979784904107},
	number = {1},
	urldate = {2017-09-21},
	journal = {J. Colloid Interface Sci.},
	author = {Chauveteau, Guy and Tirrell, M. and Omari, A.},
	year = {1984},
	pages = {41--54},
}

@article{cuenca_submicron_2013,
	title = {Submicron {Flow} of {Polymer} {Solutions}: {Slippage} {Reduction} due to {Confinement}},
	volume = {110},
	number = {10},
	journal = {PRL},
	author = {Cuenca, Amandine and Bodiguel, Hugues},
	year = {2013},
	pages = {108304},
}

@article{saerbeck2018,
	title = {Recent upgrades of the neutron reflectometer D17 at ILL},
	volume = {51},
	doi = {10.1107/S160057671800239X},
	journal = {J. Appl. Cryst.},
	author = {T. Saerbeck and R. Cubitt and A. Wildes and G. Manzin and K. H. Andersen and P. Gutfreund},
	year = {2018},
	pages = {249-256}
}

@article{degennesConformationsPolymersAttached1980,
	title = {Conformations of polymers attached to an interface},
	volume = {13},
	issn = {0028-0836, 1476-4687},
	doi = {10.1021/ma60077a009},
	journal = {Macromol.},
	author = {{P.-G.} de Gennes},
	year = {1980},
	pages = {1069-1075}
}

@book{rubinsteinPolymerPhysics,
 title = {Polymer physics},
 author = {M. Rubinstein and R. H. Colby},
 year = {2003},
 publisher = {Oxford University Press},
 pages={177}
}

@article{degennesPolymerSolutionsInterface1981,
  title = {Polymer solutions near an interface. Adsorption and depletion layers},
  author = {de Gennes, P.-G.},
  year = {1981},
  journal = {Macromol.},
  doi = {10.1021/ma50007a007},
  volume = {14},
  number = {6},
  pages = {1637-1644}
}

@article{paridaAdsorptionOrganicMolecules2006,
  title = {Adsorption of organic molecules on silica surface},
  author = {S. K. Parida and S. Dash and S. Patel and B. K. Mishra},
  year = {2006},
  journal = {Adv. Colloid Interface Sci.},
  doi = {10.1016/j.cis.2006.05.028},
  volume = {121},
  number = {1-3},
  pages = {77-110}
}

@article{meyerInteractionsAromaticRings2003,
  title = {Interactions with Aromatic Rings in Chemical and Biological Recognition},
  author = {E. A. Meyer and R. K. Castellano and F. Diederich},
  year = {2003},
  journal = {Angew. Chem. Int. Ed.},
  doi = {10.1002/anie.200390319},
  volume = {42},
  number = {11},
  pages = {1210-1250}
}

@article{leeDirectMeasurementsPolymer1991,
  title = {Direct measurements of polymer depletion layers by neutron reflectivity},
  author = {L. Lee and O. Guiselin and A. Lapp and B. Farnoux and J. Penfold},
  year = {1991},
  journal = {PRL},
  doi = {10.1103/PhysRevLett.67.2838},
  volume = {67},
  number = {20},
  pages = {2838-2841}
}

@article{gvaramiaDepletionPolyelectrolytesLikeCharged2022,
  title = {Depletion of Polyelectrolytes near Like-Charged Substrates Probed by Optical Reflectivity},
  author = {M. Gvaramia and P. Maroni and D. Kosior},
  year = {2022},
  journal = {J. Chem. Phys.},
  doi = {10.1021/acs.jpcc.2c03698},
  volume = {126},
  number = {29},
  pages = {12313-12317}
}

@article{barraudLargeSlippageDepletion2019,
    author = {Barraud, Chloé and Cross, Benjamin and Picard, Cyril and Restagno, Fréderic and Léger, Liliane and Charlaix, Elisabeth},
    title  ={Large slippage and depletion layer at the polyelectrolyte/solid interface},
    journal  = {Soft Matter},
    year  = {2019},
    volume  = {15},
    issue  = {31},
    pages  = {6308-6317},
    publisher  = {The Royal Society of Chemistry},
    doi  = {10.1039/C9SM00910H},
    url  = {http://dx.doi.org/10.1039/C9SM00910H},
    }

@article{depabloHydrodynamicChangesDepletion1992,
author = {De Pablo, Juan José and Öttinger, Hans Christian and Rabin, Yitzhak},
title = {Hydrodynamic changes of the depletion layer of dilute polymer solutions near a wall},
journal = {AIChE Journal},
volume = {38},
number = {2},
pages = {273-283},
doi = {https://doi.org/10.1002/aic.690380213},
year = {1992}
}

@article{ausserreHydrodynamicThickeningDepletion1991,
  title = {Hydrodynamic Thickening of Depletion Layers in Colloidal Solutions},
  author = {D. Ausserré and J. Edwards and J. Lecourtier and H. Hervet and F. Rondelez},
  year = {1991},
  journal = {EPL},
  doi = {10.1209/0295-5075/14/1/006},
  volume = {14},
  number = {1},
  pages = {33-38}
}

@article{merleHighFieldNMR2022,
  title = {High‐Field NMR, Reactivity, and DFT Modeling Reveal the $\gamma‐Al_{2}O_{3}$ Surface Hydroxyl Network},
  author = {Nicolas Merle and Tarnuma Tabassum and Susannah L. Scott and Allesandro Motta and Kai Szeto and Mostafa Taoufik and Régis Michaël Gauvin and Laurent Delevoye},
  year = {2022},
  journal = {Angew. Chem.},
  pages = {e202207316},
  doi = {10.1002/ange.202207316},
  volume = {134}
}

@article{shenSurfactinStructuresInterfaces2011,
  title = {Surfactin Structures at Interfaces and in Solution: The Effect of pH and Cations},
  author = {Hsin-Hui Shen and Tsung-Wu Lin and Robert K. Thomas and Diana J. F. Taylor and Jeffrey Penfold},
  year = {2011},
  journal = {J. Phys. Chem. B},
  doi = {10.1021/jp109360h},
  volume = {115},
  number = {15},
  pages = {4427-4435}
}

@article{Tatek2011,
  title = {Structural properties of atactic polystyrene adsorbed onto solid surfaces},
  author = {Yergou B. Tatek and Mesfin Tsige},
  year = {2011},
  journal = {J. Chem. Phys.},
  doi = {10.1063/1.3658046},
  volume = {135},
  pages = {174708}
}

@article{silberberg1968,
author = {Silberberg,A. },
title = {Adsorption of Flexible Macromolecules. IV. Effect of Solvent–Solute Interactions, Solute Concentration, and Molecular Weight},
journal = {J. Chem. Phys.},
volume = {48},
number = {7},
pages = {2835-2851},
year = {1968},
doi = {10.1063/1.1669540},

URL = { 
        https://doi.org/10.1063/1.1669540
    
},
eprint = { 
        https://doi.org/10.1063/1.1669540
    
}

}

@article{joanny1979,
author = {Joanny, J. F. and Leibler, L. and De Gennes, P. G.},
title = {Effects of polymer solutions on colloid stability},
journal = {J. Polym. Sci., Polym. Phys. Ed.},
volume = {17},
number = {6},
pages = {1073-1084},
doi = {https://doi.org/10.1002/pol.1979.180170615},
year = {1979}
}

@article{korolkovas2017,
  title = {Polymer Brush Collapse under Shear Flow},
  author = {A. Korolkovas and C. Rodriguez-Emmenegger and A. de los Santos Pereira and A. Chennevière and F. Restagno and M. Wolff and F. A. Adlmann and A. J. C. Dennison and P. Gutfreund},
  year = {2017},
  journal = {Macromol.},
  doi = {},
  volume = {50},
  pages = {1215-1224}
}

@article{grzelka2021,
  title = {Slip and friction mechanisms at polymer semi-dilute solutions / solid interfaces},
  author = {Marion Grzelka and Iurii Antoniuk and Eric Drockenmuller and Alexis Chennevière and Liliane Léger and Frédéric Restagno},
  year = {2021},
  journal = {Macromol.},
  doi = {},
  pages = {4910-4917},
  volume = {54},
  number = {10}
}

@article{napolitano2020,
  title = {Irreversible adsorption of polymer melts and nanoconfinement effects},
  author = {Simone Napolitano},
  year = {2020},
  journal = {Soft Matter},
  doi = {},
  volume = {16},
  number = {23},
  pages = {5348-5365}
}

@inproceedings{cousin2018neutron,
  title={Neutron reflectivity for soft matter},
  author={Cousin, Fabrice and Chennevi{\`e}re, Alexis},
  booktitle={EPJ Web of Conferences},
  volume={188},
  pages={04001},
  year={2018},
  organization={EDP Sciences}
}

@article{penfold1990application,
  title={The application of the specular reflection of neutrons to the study of surfaces and interfaces},
  author={Penfold, J and Thomas, RK},
  journal={J. Condens. Matter Phys.},
  volume={2},
  number={6},
  pages={1369},
  year={1990},
  publisher={IOP Publishing}
}

@article{russell1990x,
  title={X-ray and neutron reflectivity for the investigation of polymers},
  author={Russell, TP},
  journal={Mater. Sci. Rep.},
  volume={5},
  number={4},
  pages={171--271},
  year={1990},
  publisher={Elsevier}
}

@article{guyard2021near,
  title={Near-surface rheology and hydrodynamic boundary condition of semi-dilute polymer solutions},
  author={Guyard, Gabriel and Vilquin, Alexandre and Sanson, Nicolas and Jouenne, St{\'e}phane and Restagno, Fr{\'e}d{\'e}ric and McGraw, Joshua D},
  journal={Soft Matter},
  volume={17},
  number={14},
  pages={3765--3774},
  year={2021},
  publisher={Royal Society of Chemistry}
}

@article{nelson2019refnx,
  title={refnx: neutron and X-ray reflectometry analysis in Python},
  author={Nelson, Andrew RJ and Prescott, Stuart W},
  journal={J. Appl. Cryst.},
  volume={52},
  number={1},
  pages={193--200},
  year={2019},
  publisher={International Union of Crystallography}
}

@article{stuart1985experimental,
	title={Experimental aspects of polymer adsorption at solid/solution interfaces},
	author={Stuart, Martien A Cohen and Cosgrove, Terence and Vincent, Brian},
	journal={Adv. Colloid Interface Sci.},
	volume={24},
	pages={143--239},
	year={1985},
	publisher={Elsevier}
}

@article{kawaguchi1992polymer,
	title={Polymer adsorption at solid-liquid interfaces},
	author={Kawaguchi, Masami and Takahashi, Akira},
	journal={Adv. Colloid Interface Sci.},
	volume={37},
	number={3-4},
	pages={219--317},
	year={1992},
	publisher={Elsevier}
}

@inbook{fleer1998general,
	title={General Features of Polymers at Interfaces},
	author={Fleer, GJ and Stuart, MA Cohen and Scheutjens, JMHM and Cosgrove, T and Vincent, B and Fleer, GJ and Stuart, MA Cohen and Scheutjens, JMHM and Cosgrove, T and Vincent, B},
	journal={Polymers at Interfaces},
	pages={27--42},
	year={1998},
	publisher={Springer}
}

@article{guiselin1992irreversible,
	title={Irreversible adsorption of a concentrated polymer solution},
	author={Guiselin, OEPL},
	journal={EPL},
	volume={17},
	number={3},
	pages={225},
	year={1992},
	publisher={IOP Publishing}
}

@article{thees2020review,
	title={Review and reproducibility of forming adsorbed layers from solvent washing of melt annealed films},
	author={Thees, Michael F and McGuire, Jennifer A and Roth, Connie B},
	journal={Soft Matter},
	volume={16},
	number={23},
	pages={5366--5387},
	year={2020},
	publisher={Royal Society of Chemistry}
}

@article{morfin1999temperature,
	title={Temperature and shear rate dependence of small angle neutron scattering from semidilute polymer solutions},
	author={Morfin, I and Lindner, P and Boue, F},
	journal={Macromol.},
	volume={32},
	number={21},
	pages={7208--7223},
	year={1999},
	publisher={ACS Publications}
}

@article{brochard1992shear,
	title={Shear-dependent slippage at a polymer/solid interface},
	author={Brochard, F and De Gennes, P Gd},
	journal={Langmuir},
	volume={8},
	number={12},
	pages={3033--3037},
	year={1992},
	publisher={ACS Publications}
}

\end{document}